\documentclass[a4paper]{article}

\usepackage[margin=1in]{geometry}

\usepackage[T1]{fontenc}
\newcommand{\changefont}[3]{
\fontfamily{#1} \fontseries{#2} \fontshape{#3} \selectfont}

\changefont{ptm}{m}{n}

\usepackage{epsfig}
\usepackage{amsfonts}
\usepackage{amssymb}
\usepackage{float}
\usepackage{epstopdf}
\usepackage{graphicx}
\usepackage{mathrsfs}
\usepackage{amsmath}
\usepackage{mathtools}
\usepackage{mathrsfs}

\usepackage{setspace} 
 \usepackage{graphicx}    
  \usepackage{graphics}

\newcommand \be{\begin{equation}}
\newcommand \ee{\end{equation}}
\newcommand \ba{\begin{eqnarray}}
\newcommand \ea{\end{eqnarray}}

\def\bit{\begin{itemize}}
\def\eit{\end{itemize}}

\usepackage{amsfonts}
\usepackage{amssymb}
\usepackage{graphics}
\usepackage{mathrsfs}
\usepackage{color}
\newtheorem{remark}{Remark}[section]

\newtheorem{theorem}{Theorem}[section]

\newtheorem{lemma}{Lemma}[section]

\long\def\symbolfootnote[#1]#2{\begingroup%
\def\thefootnote{\fnsymbol{footnote}}\footnote[#1]{#2}\endgroup} 

\begin{document}

%

\begin{center}
\Large \textbf{Replication of Li-Yorke Chaos Near a Homoclinic Orbit}
\end{center}

\vspace{-0.3cm}
\begin{center}
\normalsize \textbf{Marat Akhmet$^{a,} \symbolfootnote[1]{Corresponding Author Tel.: +90 312 210 5355,  Fax: +90 312 210 2972, E-mail: marat@metu.edu.tr}$, Michal Fe\v ckan$^{b,c}$, Mehmet Onur Fen$^a$, Ardak Kashkynbayev$^a$} \\
\vspace{0.2cm}
\textit{\textbf{\footnotesize$^a$Department of Mathematics, Middle East Technical University, 06800, Ankara, Turkey}} \\
\textit{\textbf{\footnotesize$^b$Department of Mathematical Analysis and Numerical Mathematics, Comenius University in Bratislava, Mlynsk\'a dolina, 842 48 Bratislava, Slovakia}} \\
\textit{\textbf{\footnotesize$^c$Mathematical Institute of Slovak Academy of Sciences, \v Stef\'anikova 49, 814 73 Bratislava, Slovakia}}
\vspace{0.1cm}
\end{center}

\vspace{0.3cm}

\begin{center}
\textbf{Abstract}
\end{center}

\noindent\ignorespaces

    We prove the presence of chaos near a homoclinic orbit in the modified Li-Yorke sense \cite{Akh14} by implementing chaotic perturbations. A Duffing oscillator is considered to show the effectiveness of our technique, and simulations that support the theoretical results are depicted. Ott-Grebogi-Yorke and Pyragas control methods are used to stabilize almost periodic motions.

\vspace{0.2cm}
 
\noindent\ignorespaces \textbf{Keywords:} Li-Yorke chaos; Homoclinic orbit; Duffing oscillator; Almost periodic solutions

\section{Introduction}

The investigations of chaos theory for continuous-time dynamics started due to the needs of real world applications, especially with the studies of Poincar\'{e}, Cartwright and Littlewood, Levinson, Lorenz, and Ueda \cite{Cartwright1,Levinson,Lorenz63,poincare,Ueda}. Chaotic dynamics has high effectiveness in various fields such as the analysis of electrical processes of neural networks, weather phenomena, mechanical systems, optimization and self-organization problems in robotics, and brain dynamics. The reason for that is the opportunities provided by the dynamical structure of chaos.

To explain the extension procedure of our paper, let us give the following information. It is known that if one considers the evolution equation $u'=L[u]+I(t),$ where $L[u]$ is a linear operator with spectra placed out of the imaginary axis of the complex plane, then a function $I(t)$ being considered as an input with a certain property (boundedness, periodicity, almost periodicity) produces through the equation the output, a solution with a similar property, boundedness/periodicity/almost periodicity. In particular, in our paper, we solved a similar problem when the linear system has eigenvalues with negative real parts and input is considered as a chaotic set of functions with a known type. Our results are different in the sense that the input and the output are not single functions, but \textit{a collection of functions}. In other words, we prove that both the input and the output are \textit{chaos} of the same type for the discussed equation. The way of our investigation is arranged in the well accepted traditional mathematical fashion, but with a new and a more complex way of arrangement of the connections between the input and the output. The same is true for the control results discussed in the paper. If one considers an element of the chaotic set as the chaotic function, then we may consider our results through input-output mechanisms where input and output are of the same nature, that is, they are chaotic functions.

The paper \cite{Akh8}, where we discuss an extension mechanism of chaos, is about the replication of specific types of chaos, such as Devaney, Li-Yorke chaos and chaos obtained through period-doubling cascade. In this process,  we consider the generator-replicator systems such that the generator is considered as a system of the form
\begin{eqnarray} \label{generator}
x'=F(t,x),
\end{eqnarray}
where $F: \mathbb R \times \mathbb R^m \to \mathbb R^m $ is a continuous function in all its arguments
and the replicator is assumed to have the form
\begin{eqnarray} \label{replicator}
y'=\Lambda y+H(x,y),
\end{eqnarray}
where $H: \mathbb R^m \to \mathbb R^n $ is a continuous function in all its arguments and $ \Lambda $ is the $n \times n $ real valued hyperbolic matrix.

The rigorous results of the extension mechanism emphasize that  system $(\ref{replicator})$ is chaotic in the same way as system $(\ref{generator}).$ Replication of chaos through intermittency is also shown through simulations in paper \cite{Akh8}, where one can find new definitions for chaotic sets of functions, and precise descriptions for the ingredients of Devaney and Li-Yorke chaos in continuous-time dynamics, which are used as tools for the extension procedure.

In the case that the matrix $\Lambda$ represented in system (\ref{replicator}) is hyperbolic, we have not been able to find a way to insert a term nonlinear in $y$ in the system to preserve the results of paper \cite{Akh8} and, instead, we were forced to handle system $(\ref{replicator})$ with $H(x,y) \equiv H(x) $ to achieve success in the theoretical results \cite{Akh13}. In other words, we could not achieve the extension of Devaney and/or Li-Yorke chaos when nonlinearity with respect to $y$ is included in the system and the eigenvalues of the matrix $\Lambda$ are allowed to possess positive real parts as well as negative real parts. However, in the present paper, the chaos extension problem in the sense of Li-Yorke is considered for a hyperbolic matrix $\Lambda$ and with the nonlinear term of the initially non-perturbed system is also considered  for a theoretically supported chaotification process, and this is the main difference compared to \cite{Akh13,Akh8}.

Traditionally, analysis of nonlinear dynamical systems has been restricted to smooth problems, that is, smooth differential equations. Besides stability analysis of fixed points or periodic orbits, another fascinating phenomenon has been found: the existence of chaotic orbits. The presence of these orbits has the consequence that the motions of the system depend sensitively on initial conditions and the asymptotic behavior of orbits in the future is unpredictable. Such a chaotic behavior of solutions can be explained mathematically by showing the existence of a transverse homoclinic point of the time map with the corresponding invariant Smale horseshoe \cite{Guc}. In general, however, it is not easy to show the existence of a transverse homoclinic point. To this purpose, the perturbation approach, by now known as the Melnikov method, is a powerful method \cite{Fec1}-\cite{Fec3}. The starting point is a nonautonomous system, the unperturbed system/equation, with a (necessarily) nontransverse homoclinic orbit. Then it is known that, if we take the perturbed system by adding a periodic (or almost periodic) perturbation of sufficiently small amplitude to the unperturbed system and a certain Melnikov function has a simple zero at some point, the perturbed system has a transverse homoclinic point with the corresponding Smale horseshoe \cite{Fec4}.

In the present study, we will consider these systems and perturb them in a unidirectional way through exogenous chaotic forcing terms to achieve propagation of the chaotic behavior. In other words, the influence of the chaos of a system on another one will be mentioned in the paper such that as a result the latter behaves also chaotically. Chaotification of systems with asymptotically stable equilibriums through different type of perturbations can be found in \cite{Akh2,Akh4,Akh5,Akh7,Akh8,Akh12,AkhF}. Endogenously generated chaotic behavior of systems are well investigated in the literature. The systems of Lorenz \cite{Lorenz63}, R\"{o}ssler \cite{Rossler76} and Chua \cite{Chua93,Chua86} as well as the Van der Pol \cite{Cartwright1,Levi,Levinson} and Duffing \cite{Luo12,Moon,Th02} oscillators can be considered as systems which are capable of generating chaos endogenously. We will make use of Li-Yorke chaos in the extension mechanism through exogenous perturbations, and essentially indicate in the present paper that not only endogenous structure of systems, but also exogenous chaotic perturbations can give rise to observation of chaotic behavior.

Infinitely many periodic solutions can serve as a basis of Li-Yorke chaos as well as infinitely many quasi-periodic or almost periodic solutions can also serve as a basis for the developed Li-Yorke chaos \cite{Akh14}. In the present study, we follow the definition of continuous Li-Yorke chaos  modified in \cite{Akh14} and consider quasi-periodic and almost periodic solutions as a basis of Li-Yorke chaos.

In the next section, we introduce the systems which will be under investigation and give information about the properties of these systems under some conditions.

\section{Preliminaries}

Let us consider the systems
\begin{eqnarray}\label{1}
x'=F(x,t)
\end{eqnarray}
and
\begin{eqnarray}\label{e1}
z'=f(z,t)
\end{eqnarray}
where the function $F:\mathbb R \times \mathbb R^{m}\to \mathbb R^{m}$ is continuous  in all its arguments and $f: \mathbb R \times  \mathbb R^{n} \to \mathbb R^n $ is continuous, $ C^2$ function in $x$ and satisfy the following conditions:

\begin{itemize}
    \item [(\textbf{C1})] $f$ is  $1$-periodic in $t;$
    \item [(\textbf{C2})] The system (\ref{e1}) has a hyperbolic periodic solution  $p(t)$ with a homoclinic solution $q(t)$ such that the variational equation $v'=D_uf(q(t),t)v$ has the only zero solution bounded on $\mathbb R$.
\end{itemize}

To extend chaos generated by equation (\ref{1}), we perturb system (\ref{e1}) through the solutions of (\ref{1}) to achieve the system
\begin{eqnarray}\label{2}
u'=f(u,t)+h(x),
\end{eqnarray}
where $h:\mathbb R^{m} \to \mathbb R^{n}$ is a continuous function. The system \eqref{1} has infinitely many periodic solutions as well as the system \eqref{e1} around a homoclinic orbit $ q(t) $. If the fundamental periods of the systems \eqref{1} and \eqref{e1} are commensurable, that is the quotient of the fundamental periods are rational, then the system \eqref{2} has periodic solutions. Otherwise, if the fundamental periods of the systems \eqref{1} and \eqref{e1} are incommensurable, the system \eqref{2} has quasi-periodic or more general almost periodic solutions. Thus, in this study Li-Yorke chaos is generated with infinitely many almost periodic motions in basis instead of periodic motions. 

Throughout the paper, we will make use of the usual Euclidean norm for vectors and the norm induced by the Euclidean norm for matrices \cite{Horn92}. 

The following conditions are required:
\begin{enumerate}
    \item[\bf (C3)] There exists a positive  number $H_0$ such that $\displaystyle \sup_{x\in \mathbb R^m, t\in \mathbb R} \left\|F(x,t)\right\|\le H_0;$
    \item[\bf (C4)] There exist positive numbers $L_1,$ $L_2$ and $L_3$ such that  
    $$ L_1\left\|x_1-x_2\right\| \leq \left\|h(x_1)-h(x_2)\right\| \leq L_2\left\|x_1-x_2\right\|$$ for all $x_1,$ $x_2 \in \mathbb R^{m},$ and  $$\left\|f(u_1,t)-f(u_2,t))\right\| \leq L_3\left\|u_1-u_2\right\|$$ for all  $u_1,u_2 \in \mathbb R^n,$ $t\in \mathbb R.$ 
\end{enumerate}

It is worth noting that the results of our study are also true even if we replace the non-autonomous system (\ref{1}) by the autonomous equation
\begin{eqnarray}\label{autonomus_master}
    x'=\overline{F}(x),
\end{eqnarray}
where the function $\overline{F}:\mathbb R^{m}\to \mathbb R^{m}$  is continuous with the condition which is a counterpart of $(C3)$.

Under the  condition (C2), it is known that $q(t)$ is a transversal homoclinic
orbit, i.e., taking the $1$-time map $G : \mathbb R^m\to \mathbb
R^m$ of the system (\ref{e1}), it has a hyperbolic fixed point
$p(0)$ with a transversal homoclinic orbit $\left\lbrace
q(n)\right\rbrace_{n\in\mathbb Z}$. Following Sections 3 and 4 of
\cite{P1}, especially Theorem 4.8 of \cite{P1}, we 
get a collection of bounded solutions $\left\lbrace
\nu_\beta(t)\right\rbrace _{\beta\in S_m}$ of system
(\ref{e1}) orbitally near $q(t)$, where the index set $S_m,$ $m
\ge 2,$ is the set of doubly infinite sequences
$
a=\big(\ldots,a_{-1},a_0,a_1,\ldots\big)
$
with $a_i \in \{ 1,\ldots,m\}$  for all $i\in \mathbb{Z},$ i.e., 
$
S_m= \{1,2,\ldots,m\}^{\mathbb{Z}}
$ such that each linear
system
\begin{equation}\label{e2}
    v'=D_uf(\nu_\beta(t),t)v
 \end{equation}
has an exponential dichotomy on $\mathbb R$ with uniform positive constants $K$ and $\alpha$ and projections $P_\beta$:
 \begin{equation}\label{e5}
    \begin{gathered}
        \big\|V_\beta(t)P_\beta V_\beta(s)^{-1} \big\|\le Ke^{-\alpha(t-s)} \, \forall t,s, \, t\ge s,\\
        \big\|V_\beta(t)(I- P_\beta)V_\beta(s)^{-1}\big\|\le Ke^{\alpha(t-s)}\, \forall t,s,\, t\le s,
    \end{gathered}
 \end{equation}
where $V_\beta$ is the fundamental solution of the system
(\ref{e2}).

By Theorem 4.8 of \cite{P1}, an iterative $G^{\bar l}$, for some fixed $\bar
l\in\mathbb N$, is conjugate to the Bernoulli shift on an
invariant compact subset $\textit{I}\subset\mathbb{R}^n$, $G^{\bar
l} : \textit{I}\to \textit{I}$. So $G^{\bar l}$ has $i$-periodic
orbits in $\textit{I}$ for any natural number $i$. This gives that the
original map $G$ has periodic orbits with periods $i\bar l$ starting in 
$\textit{I}$. Since by definition $\nu_\beta(0)=\varsigma_\beta$ for some
$\varsigma_\beta\in\textit{I}$ and then $G^k(\varsigma_\beta)=\nu_\beta(k)$ for any $k\in\mathbb Z$, we see that among
these $\nu_\beta(t)$ there are $i\bar l$-periodic solutions for
any $i\in\mathbb N$.

Introducing the new variable $ y $ through $u=y+\nu_\beta,$ 
system (\ref{2}) can be written in the following form:
\begin{eqnarray}\label{e3}
y'=D_uf(\nu_\beta,t)y + f(y+\nu_\beta,t)-f(\nu_\beta,t)  -D_uf(\nu_\beta,t)y+h(x).
\end{eqnarray}
 Since $ f $ is  $ C^2 $ function, there exist positive numbers $ P_1 $ and $ P_2 $ such that $\displaystyle \sup_{t\in \mathbb R, \beta\in S_m} \left\|D_uf(\nu_\beta(t),t)\right\|\le P_1$ and $\displaystyle \sup_{t\in \mathbb R, \beta\in S_m} \left\|D_{uu}f(\nu_\beta(t),t)\right\|\le P_2$ for each bounded solution $ \nu_\beta(t) $ of (\ref{e1}).

Our main assumption is the existence of a nonempty set $\mathscr{A}_x$ of all solutions of system (\ref{1}), uniformly bounded on $\mathbb R.$ That is, there exists a positive real number $H$ such that $\displaystyle \sup_{t \in \mathbb R} \left\|x(t)\right\| \leq H$ for all $x(t) \in \mathscr{A}_x.$

The following conditions are also needed:
\begin{itemize}
   \item[\bf (C5)] There exists a positive number $M_f$  such that $\displaystyle \sup_{ u\in \mathbb R^n, t\in \mathbb R} \left\|f(u,t)\right\|\le M_f;$
   \item[\bf (C6)] $M_h < \displaystyle \frac{\alpha^2}{16K^2P_2^2},$  where $M_h = \displaystyle \sup_{\left\| x \right\| \le H} \left\|h(x) \right\|.$
\end{itemize}
Using the dichotomy theory \cite{Cop}, one can verify that for a given solution $x(t)$ of system (\ref{1}), a bounded on $\mathbb R$ function $ \phi_{x(t)}^{\beta}(t) = y(t) $ is a solution of system (\ref{e3}) if and only if the following integral equation is satisfied
\begin{eqnarray}
\label{e6}
 y(t)=\int\limits_{-\infty}^{\infty}G_\beta(t,s)(H_1(y(s),\nu_\beta(s),s)+h(x(s)))ds,
\end{eqnarray}
where
\begin{eqnarray*}
G_\beta(t,s)=\left\{\begin{array}{ll}V_\beta(t)P_\beta V_\beta(s)^{-1}&\quad t\ge s,\\ V_\beta(t)(I-P_\beta) V_\beta(s)^{-1}&\quad t\le s,\end{array}\right.
\end{eqnarray*}
and
\begin{eqnarray*}
H_1(y_1,y_2,s)=f(y_1+y_2,s)-f(y_2,s)-D_uf(y_2,s)y_1.
\end{eqnarray*}

Now, we are ready to prove the following result.

\begin{lemma}\label{lem11} If conditions $ (C1)-(C6) $ hold, then for each $ x(t) \in \mathscr{A}_x $ there exists a unique solution $\phi_{x(t)}^\beta(t)$ of (\ref{e3}) such that $ \displaystyle \sup_{t \in \mathbb R} \left\|\phi_{x(t)}^\beta(t)\right\| \leq r_1$ for a constant
$
r_1=\displaystyle \frac{\frac{4KP_2}{\alpha}M_h}{1+\sqrt{1-\frac{16K^2P_2^2}{\alpha^2}M_h}}.
$
\end{lemma}
\noindent \textbf{Proof.} Consider the set $ C_0(\mathbb R) $ of continuous functions $ y(t) $ satisfying $ \left\|y\right\|_{0}  \leq r_1,$ where $\left\|y\right\|_0 = \displaystyle \sup_{t \in \mathbb R} \left\|y(t)\right\|.$  Define the operator $ \Pi $ on $ C_0 $ as
\begin{eqnarray*}
\Pi y(t)=\int\limits_{-\infty}^{\infty}G_\beta(t,s)(H_1(y(s),\nu_\beta(s),s)+h(x(s)))ds.
\end{eqnarray*}
Using the mean value theorem we obtain  that
\begin{eqnarray*}
& \|H_1(y(s),\nu_\beta(s),s)\| & \le \int_0^1\left\|D_uf(\theta y_1+y_2,s)-D_uf(y_2,s)\right\|d\theta  \|y_1\| \\
&& \le \int_0^1\int_0^1\left\|D_{uu}f(\tau\theta y_1+y_2,s)\right\|d\tau d\theta  \|y_1\|^2 \\ 
&& \le P_2\|y_1\|^2.
\end{eqnarray*}
Hence, the inequality
$$ \left\|\Pi y(t)\right\| \le  \frac{2K(P_2\|y\|_0^2+M_h)}{\alpha}
$$
is valid so that
$$
\left\|\Pi y\right\|_0\le\frac{2K(P_2\|y\|_0^2+M_h)}{\alpha}.
$$
Let 
\begin{eqnarray*}
H_2(y_1,y_2,y_3,s)=f(y_1+y_2,s)-f(y_3+y_2,s) -D_uf(y_2,s)(y_1-y_3).
\end{eqnarray*}
Similarly, we derive by the mean value theorem
\begin{eqnarray*} 
&&\|H_2(y_1,y_2,y_3,s)\| \le \int_0^1\big\|D_uf(\theta y_1+(1-\theta)y_3+y_2,s)  -D_uf(y_2,s)\big\|d\theta  \|y_1-y_3\| \\
&&\le \int_0^1\int_0^1\left\|D_{uu}f(\tau(\theta y_1+(1-\theta)y_3)+y_2,s)\right\|d\tau     (\theta \|y_1\|+(1-\theta)\|y_3\|)d\theta \|y_1-y_3\|\\
&&\le\int_0^1\int_0^1\left\|D_{uu}f(\tau(\theta y_1+(1-\theta)y_3)+y_2,s)\right\|d\tau d\theta   \max\{\|y_1\|,\|y_3\|\}\|y_1-y_3\|\\
&& \le P_2\max\{\|y_1\|,\|y_3\|\}\|y_1-y_3\|.
\end{eqnarray*}
Then one can confirm that
\begin{eqnarray*}
\left\|\Pi y_1(t)-\Pi y_2(t) \right\|  \le \frac{2KP_2}{\alpha}\max\{\|y_1\|_0,\|y_2\|_0\}\left\|y_1-y_2\right\|_0 .
\end{eqnarray*}
Hence, we arrive at
$$
\left\|\Pi y_1-\Pi y_2 \right\| _0 \le \frac{2KP_2}{\alpha}\max\{\|y_1\|_0,\|y_2\|_0\}\left\|y_1-y_2\right\|_0.
$$
Next, the quadratic function  $$q(r)=\displaystyle \frac{2KP_2}{\alpha}r^2-r+\frac{2KP_2}{\alpha}M_h$$ has two positive roots $0<r_1<r_2$ when
$$
M_h<\displaystyle \frac{\alpha^2}{16K^2P_2^2},
$$
which is satisfied by (C6). Moreover, $q'(r_1)<0$, i.e., $\displaystyle \frac{2KP_2}{\alpha}r_1<\frac{1}{2}$. This means that for the ball $B_{r_1}=\{y\in C_0(\mathbb R) : \|y\|_0\le r_1\}$, the map $\Pi : B_{r_1}\to B_{r_1}$ is a contraction with coefficient $1/2$. Note that
$
r_1  \displaystyle \le \frac{4KP_2}{\alpha}M_h,
$
i.e., the smaller $M_h$ the smaller $B_{r_1}$, as we can expect this. Thus, (\ref{e3}) admits a unique solution from $  C_0(\mathbb R). $ This finalizes the proof of the lemma.  $\square$

\subsection{Almost Periodic Functions}
A continuous  function $ \mathcal{F}(t) $ is said to be almost periodic, if for any $ \epsilon > 0 $ there exists $ l > 0 $ such that for any interval with length $ l $ there exists a number $ \omega $ in this interval satisfying $ \left\|\mathcal{F}(t+\omega)-\mathcal{F}(t) \right\| < \epsilon $ for all $ t \in \mathbb{R }$ \cite{Hale80,Lev82,SP}.
Now, let us show that the bounded solution is almost periodic.
\begin{lemma}\label{almost}
If in addition in Lemma \ref{lem11}, $ x(t) $ and $\nu_\beta(t)$
are almost periodic then $\phi_{x(t)}^\beta(t)$ of \eqref{e3} is
almost periodic as well. Moreover, when $x(t)$ and $\nu_\beta(t)$
are both $i\bar l$-periodic for some $i\in\mathbb{N}$, then
$\phi_{x(t)}^\beta(t)$ is also $i\bar l$-periodic.
\end{lemma}

\noindent \textbf{Proof.}
Consider the set $ B_{r_1}^{AP} $ of all almost periodic functions of $B_{r_1}$. By \eqref{e3}, the function $z(t)=\Pi y(t)$ from the above proof is defined as the bounded solution of
\begin{equation}\label{e3b}
z'=A_\beta(t)z+ f_\beta(t),
\end{equation}
where
$$
A_\beta(t)=D_uf(\nu_\beta(t),t) 
$$
and
$$
f_\beta(t)=f(y(t)+\nu_\beta(t),t)-f(\nu_\beta(t),t) -D_uf(\nu_\beta(t),t)y(t)+h(x(t)).
$$
If $y(t)$ is almost periodic, by \cite{Hale80}, $A_\beta(t)$ and $f_\beta(t)$ are almost periodic. Then using results of \cite[pp. 72]{Cop}, we know that $z(t)$ is also almost periodic. This means that $ \Pi : B_{r_1}^{AP}\to B_{r_1}^{AP}$. Since $ \Pi $ is contractive, its fixed point $\phi_{x(t)}^\beta(t)$ is almost periodic.

Next, when $x(t)$ and $\nu_\beta(t)$ are both $i\bar l$-periodic
for some $i\in\mathbb{N}$, then
$\overline{y}(t)=\phi_{x(t)}^\beta(t+i\bar l)$ satisfies
\eqref{e3} and $\|\overline{y}\|\le r_1$. The uniqueness of such a
solution implies $\overline{y}(t)=y(t)$, so $\phi_{x(t)}^\beta(t)$
is also $i\bar l$-periodic. The proof is finished.
$\square$

\subsection{Li-Yorke Chaos}
In the original paper of Li and Yorke \cite{Li75}, infinitely many periodic solutions, which are separated from the elements of a scrambled set, is introduced. In the present study, we make use of the definition defined in \cite{Akh14}, where Akhmet et al. modified the definition of Li-Yorke chaos by considering infinitely many almost periodic solutions as a basis of the developed continuous Li-Yorke chaos instead of periodic solutions, which are separated from the elements of the scrambled set, to generate the Li-Yorke chaos.

Let us denote by
\begin{eqnarray} \label{collection}
\mathscr{B}=\left\{\psi(t)~|~ \psi: \mathbb R \to K ~\textrm{is}  ~\textrm{continuous}  \right\}
\end{eqnarray}
a collection of functions, where $K \subset \mathbb R^{p}$  is a bounded region.
Since the concept of  \textit{chaotic set of functions} is used in the theoretical discussions, let us explain briefly the ingredients of Li-Yorke chaos for the set $\mathscr{B},$ which are introduced in paper \cite{Akh14}. The proofs indicated in Section \ref{main_lemmas} are predicated on the definitions of these ingredients. For more information about Devaney and Li-Yorke chaos, one can see \cite{Akin03,Ciklova06,Dev90,Kloeden06,Li75,Palmer2000}.

Let us introduce the following ingredients of Li-Yorke chaos for the set $\mathscr{B}.$

\begin{enumerate}
\item[ (LY1)] A couple of functions $ \left( \psi(t), \overline{\psi}(t) \right) \in \mathscr{B} \times \mathscr{B}$ is called proximal if for arbitrary small $\epsilon>0$ and arbitrary large $E>0,$ there exists an interval $ J $ of length not less than $E$ such that $\left\|\psi(t)-\overline{\psi}(t)\right\| < \epsilon,$ for each $t\in J$;
\item[ (LY2)] A couple of functions $\left( \psi(t), \overline{\psi}(t) \right) \in \mathscr{B} \times \mathscr{B}$ is frequently $(\epsilon_0, \Delta)-$separated if there exist positive real numbers $\epsilon_0, \Delta$ and infinitely many disjoint intervals of length not less than $\Delta$, such that $\left\|\psi(t)-\overline{\psi}(t)\right\| > \epsilon_0,$ for each $t$ from these intervals.
\end{enumerate}

A couple of functions $\left( \psi(t), \overline{\psi}(t) \right) \in \mathscr{B} \times \mathscr{B}$ is a Li$-$Yorke pair if it is proximal and frequently $(\epsilon_0, \Delta)$-separated for some positive numbers $\epsilon_0$ and $\Delta.$ On the other hand, a set $\mathscr{C} \subset \mathscr{B}$ is called a scrambled set if $\mathscr{C}$ does not contain any almost periodic function and each couple of different functions inside $\mathscr{C} \times \mathscr{C}$ is a Li$-$Yorke pair.

$\mathscr{B}$ is called a Li$-$Yorke chaotic set if: $ (i) $ it admits a countably infinite set of almost periodic functions; $ (ii) $ there exists an uncountable scrambled subset $ \mathscr{C}\subset \mathscr{B};$   $ (iii) $ for any $\psi(t)\in \mathscr{C}$ and any almost periodic $\overline{\psi}(t)\in \mathscr{B},$ the pair $\left(\psi(t),\overline{\psi}(t)\right)$ is frequently $(\epsilon_0, \Delta)-$separated for some positive real numbers $\epsilon_0$ and $\Delta.$

Let us introduce the sets of functions
\begin{eqnarray}
\begin{array}{l} \label{A_yb}
\mathscr{A}_y^{\beta}=\left\{\phi_{x(t)}^{\beta}(t) ~|~  x(t) \in \mathscr{A}_x \right\}, \ \beta \in S_m.
\end{array}
\end{eqnarray}

The next section is devoted for the clarification of the  theoretical results for the chaos extension in systems of the form $(\ref{1})+(\ref{2}).$

\section{Extension of Chaos}\label{main_lemmas}

The present section is devoted for the rigorous proofs for the extension of chaos in the sense of Li$-$Yorke. We start our discussions with the first ingredient, proximality, of Li$-$Yorke chaos.

\begin{lemma} \label{proximality_proof}
Suppose that conditions $(C1)-(C6)$ are valid. If a couple $\left( x(t), \tilde{x}(t) \right) \in \mathscr{A}_x \times \mathscr{A}_x$ is proximal, then the same is true for the couple
$
\left( \phi_{x(t)}^{\beta}(t),\phi_{\tilde{x}(t)}^{\beta}(t) \right)\in \mathscr{A}_y^{\beta} \times \mathscr{A}_y^{\beta}
$
with the corresponding interval $J$ uniform for all $\beta\in S_m$.
\end{lemma}
\noindent \textbf{Proof.} Fix an arbitrary small number $ \epsilon >0 $ and an arbitrary large number $ E>0.$ Take $\epsilon_1>0$ and $E_1>0$ that will be specified later. Then there is a couple of functions $\left( x(t), \tilde{x}(t) \right) \in \mathscr{A}_x \times \mathscr{A}_x$ which is proximal with constants $\epsilon_1$, $E_1$ and interval $J_1=[a_1,a_1+3E_1]$ for some $a_1$. Let ${y}(t)=\phi_{{x}(t)}^{\beta}(t)$ and $\tilde{y}(t)=\phi_{\tilde{x}(t)}^{\beta}(t)$. Then for any $t\in [a_1,a_1+3E_1]$, it can be verified that
\begin{eqnarray*}
&y(t)-\tilde{y}(t)&=\int\limits\limits\limits_{-\infty}^{a_1} G_\beta(t,s)\Big(H_2(y(s),\nu_\beta(s),\tilde{y}(s),s) +\left( h(x(s))-h(\tilde{x}(s))\right)\Big)ds\\
&&+\int\limits\limits\limits_{a_1}^{a_1+3E_1} G_\beta(t,s)\Big(H_2(y(s),\nu_\beta(s),\tilde{y}(s),s)  +\left( h(x(s))-h(\tilde{x}(s))\right)\Big)ds\\
&&+\int\limits\limits\limits_{a_1+3E_1}^{\infty} G_\beta(t,s)\Big(H_2(y(s),\nu_\beta(s),\tilde{y}(s),s) +\left( h(x(s))-h(\tilde{x}(s))\right)\Big)ds.
\end{eqnarray*}
Using the estimates from the proof of Lemma \ref{lem11}, one can confirm that
\begin{eqnarray*}
&\left\|y(t)-\tilde{y}(t)\right\|&  \le  \Big(2P_2r_1^2+2L_2H\Big)\frac{K}{\alpha}e^{-\alpha(t-a_1)} +\Big(2P_2r_1^2+2L_2H\Big)\frac{K}{\alpha}e^{-\alpha(a_1+3E_1-t)}\\
&& +K\int\limits\limits\limits_{a_1}^{a_1+3E_1}e^{-\alpha|t-s|}\Big(P_2r_1\|y(s)-\tilde{y}(s)\|+L_2\epsilon_1\Big)ds\\
&&\le\Big(2P_2r_1^2+2L_2H\Big)\frac{K}{\alpha}\left(e^{-\alpha(t-a_1)}+e^{-\alpha(a_1+3E_1-t)}\right) 
 +\frac{2L_2K}{\alpha}\epsilon_1  \\
 &&  +KP_2r_1\int\limits\limits\limits_{a_1}^{a_1+E_1}e^{-\alpha|t-s|}\|y(s)-\tilde{y}(s)\|ds.
\end{eqnarray*}
Thus, the difference $w(t)=\|y(t)-\tilde{y}(t)\|$ satisfies the inequality
\begin{eqnarray}
&w(t)&\le \frac{K}{\alpha}\Big(2P_2r_1^2+2L_2H\Big)\Big(e^{-\alpha(t-a_1)} +e^{-\alpha(a_1+3E_1-t)}\Big)   \nonumber \\
&&  +\frac{2L_2K}{\alpha}\epsilon_1   
 +KP_2r_1\int\limits\limits\limits_{a_1}^{a_1+3E_1}e^{-\alpha|t-s|}w(s)ds
\end{eqnarray}
for $t \in [a_1,a_1+3E_1]$. Since
$
2KP_2r_1<\displaystyle \frac{\alpha}{2}<\alpha,
$
Theorem \ref{gr} can be applied to derive
$$
\|y(t)-\tilde{y}(t)\|\le \kappa\left(e^{-\delta(t-a_1)}+e^{-\delta(a_1+3E_1-t)}
+\epsilon_1\right)
$$
on $[a_1,a_1+3E_1]$, for a positive constant $\kappa$ independent of $y(t)$, $\tilde{y}(t)$, $\epsilon_1$, $a_1$, $E_1$. Hence,
$$
\|y(t)-\tilde{y}(t)\|\le \kappa\left(2e^{-\delta E_1}+\epsilon_1\right)
$$
on $[a_1+E_1,a_1+2E_1]$. Taking
$ 
\epsilon_1=\displaystyle \frac{\epsilon}{4\kappa}$ and $E_1=\displaystyle \max\left\{-\frac{\ln\frac{\epsilon}{4\kappa}}{\delta},E\right\},
$ 
we obtain
$$
\|y(t)-\tilde{y}(t)\|\le \frac{3}{4}\epsilon<\epsilon
$$
on an interval $J=[a_1+E_1,a_1+2E_1]$ with the length $E_1\ge E$. The proof is finished. $\square$

We shall continue in the next ingredient of Li-Yorke chaos in the following lemma.

\begin{lemma} \label{seperation_proof}
Suppose that conditions $(C1)-(C6)$ are valid. If a couple  $\left( x(t), \overline{x}(t) \right) \in \mathscr{A}_x \times \mathscr{A}_x$ is frequently $ (\epsilon_{0},\Delta)-$  separated for some positive numbers $\epsilon_{0}$ and $ \Delta $, then the couple 
$
\left( \phi_{x(t)}^{\beta}(t), \phi_{\overline{x}(t)}^{\beta}(t) \right) \in \mathscr{A}_y^{\beta} \times \mathscr{A}_y^{\beta}
$
is frequently $(\epsilon_{1},\overline{\Delta})-$ separated for some positive numbers $\epsilon_{1}$ and $ \overline{\Delta} $ uniformly for $\beta\in S_m$ (see \eqref{est1} and \eqref{est2}).
\end{lemma}
\noindent \textbf{Proof.}
Let $\left( x(t), \overline{x}(t) \right) \in \mathscr{A}_x \times
\mathscr{A}_x$ be a couple of functions frequently $
\left(\epsilon_0, \Delta \right)- $ separated for some $
\epsilon_0>0 $ and $ \Delta>0 $. In that case, one can find a
sequence $ \{\Delta_l\},$ satisfying $\Delta_l \geq \Delta, l \in
\mathbb{N},$ and a sequence  $ d_l \in \mathbb{R} $ such that  for
each $ l \in \mathbb{N} $ the inequality  $\lVert
x(t)-\overline{x}(t) \rVert > \epsilon_0 $ is satisfied for  $ t
\in J_{l} = [d_l,d_l+\Delta_l]$ and $ J_l \cap J_m = \emptyset $
whenever $ l \ne m. $ Set $y(t)=\phi_{x(t)}^{\beta}(t)$ and
$\overline{y}(t)=\phi_{\overline{x}(t)}^{\beta}(t)$. Now, let the
sequence $\left\lbrace b_l\right\rbrace $ be the midpoints of the
intervals $J_l,$ that is $ b_l =d_l+\Delta_l/2 $ for each $ l \in
\mathbb{N}$, and consider $J_l^1=[b_l,b_l+\Delta_l/4]\subset J_l$.

Next, for any $t\in J_l^1$, we have that
\begin{eqnarray*}
& y(t)-\overline{y}(t) &= (y(b_l)-\overline{y}(b_l))  
  + \displaystyle \int\limits^{t}_{b_l}\left(f(y(s)+\nu_\beta(s),s)-f(\overline{y}(s)+\nu_\beta(s),s)\right)ds\\
&& + \displaystyle\int\limits^{t}_{b_l}\left(h(x(s))-h(\overline{x}(s))\right)ds,
\end{eqnarray*}
which implies
\begin{eqnarray*}
&& \max_{t\in J_l^1}\|y(t)-\overline{y}(t)\| \ge \|y(b_l+\Delta_l/4)-\overline{y}(b_l+\Delta_l/4)\|\\
&&\ge \frac{L_1\Delta_l\epsilon_0}{4}-\|y(b_l)-\overline{y}(b_l)\|  
  -\left(P_1+P_2r_1\right)\frac{\Delta_l}{4}\max_{t\in J_l^1}\|y(t)-\overline{y}(t)\|,
\end{eqnarray*}
since
\begin{eqnarray*}
&& \|f(y(s)+\nu_\beta(s),s)-f(\overline{y}(s)+\nu_\beta(s),s)\| \\
&&\le \left(P_1+P_2\max\{\|y(s)\|,\|\overline{y}(s)\|\}\right)\|y(s)-\overline{y}(s)\|\\
&&\le \left(P_1+P_2r_1\right)\max_{t\in J_l^1}\|y(t)-\overline{y}(t)\|
\end{eqnarray*}
for any $s\in J_l^1$. Consequently, we obtain that
\begin{eqnarray*}
 \max_{t\in J_l^1}\|y(t)-\overline{y}(t)\|\ge \frac{L_1\Delta_l}{4}\epsilon_0 
 -\left(1+(P_1+P_2r_1)\frac{\Delta_l}{4}\right)\max_{t\in J_l^1}\|y(t)-\overline{y}(t)\|,
\end{eqnarray*}
which implies
\begin{equation}\label{ee1}
\max_{t\in J_l^1}\|y(t)-\overline{y}(t)\|\ge \frac{L_1\Delta_l\epsilon_0}{8+(P_1+P_2r_1)\Delta_l}.
\end{equation}
A real valued function $\left\|y(t)-\overline{y}(t)\right\|$ takes its maximum on the interval $J_l^1$ at $\eta_l,$ that is,
\begin{eqnarray*}
\displaystyle \max_{t \in J_l^1} \left\|y(t)-\overline{y}(t)\right\| = \left\|y(\eta_l)-\overline{y}(\eta_l)\right\|
\end{eqnarray*}
for some $\eta_l \in J_l^1$.
Using \eqref{e3}, we obtain
\begin{equation}\label{ess1}
\|y'\|_0+\|\overline{y}'\|_0\le2\left((P_1+P_2r_1)r_1+L_2H+\|h(0)\|\right).
\end{equation}
Hence, for $t\in[\eta_l,\eta_l+\overline{\Delta}_l]\subset J_l$ with $0<\overline{\Delta}_l\le\Delta_l/4$, we derive
\begin{eqnarray*}
&\left\|y(t)-\overline{y}(t)\right\| &\ge \left\|y(\eta_l)-\overline{y}(\eta_l)\right\|-\left\|\overline{y}(\eta_l)-\overline{y}(t)\right\|
-\left\|y(\eta_l)-y(t)\right\| \\ 
&&\ge \frac{L_1\Delta_l\epsilon_0}{8+(P_1+P_2r_1)\Delta_l} 
 -2\left((P_1+P_2r_1)r_1+L_2H+\|h(0)\|\right)\overline{\Delta}_l.
\end{eqnarray*}
Consequently, by taking
\begin{eqnarray*}
\overline{\Delta}_l=\min\left\{\Delta_l/4,\frac{L_1\Delta_l\epsilon_0}{4\left(8+(P_1+P_2r_1)\Delta_l\right)P_3}\right\}, 
\end{eqnarray*} where
\begin{eqnarray*}
 P_3=(P_1+P_2r_1)r_1+L_2H+\|h(0)\|,
\end{eqnarray*}   
one can attain that
\begin{eqnarray*}
 \left\|y(t)-\overline{y}(t)\right\| \ge \frac{L_1\Delta_l\epsilon_0}{2\left(8+(P_1+P_2r_1)\Delta_l\right)} 
 \ge \frac{L_1\Delta\epsilon_0}{2\left(8+(P_1+P_2r_1)\Delta\right)}
\end{eqnarray*}
for any $t\in[\eta_l,\eta_l+\overline{\Delta}_l]\subset J_l$. Note that
$$
\overline{\Delta}_l\ge\min\left\{\Delta/4,\frac{L_1\Delta\epsilon_0}{4\left(8+(P_1+P_2r_1)\Delta\right)P_3}\right\}.
$$
So, by taking
\begin{eqnarray}\label{est1}
 \epsilon_1=\frac{L_1\Delta\epsilon_0}{2\left(8+(P_1+P_2r_1)\Delta\right)},
 \end{eqnarray}
 and 
 \begin{eqnarray*} \label{est2}
 \overline{\Delta} =\min\left\{\Delta/4,\frac{L_1\Delta\epsilon_0}{4\left(8+(P_1+P_2r_1)\Delta\right)P_3}\right\}, 
\end{eqnarray*}
the proof is finished.
$\square$

Now, we state and prove the main theorem of the present section by indicating the extension of chaos in the sense of Li-Yorke for the system (\ref{1})+(\ref{2}).

\begin{theorem}\label{li-yorke_theorem}
Let $\nu_\beta(t)$ be almost periodic. If the generator \eqref{1} admits a Li-Yorke chaotic set $\mathscr{A}_x$, then the replicator \eqref{2} has  Li-Yorke chaotic sets $ \mathscr{A}_y^{\beta}, $ $\beta \in S_m$.
\end{theorem}

\noindent \textbf{Proof.}
Assume that the set $\mathscr{A}_x$ is Li-Yorke chaotic. We need to show that $\mathscr{A}_y^{\beta}$ is a Li-Yorke chaotic set as well.

First, there is a
countably infinite subset $\mathscr{P}_x\subset\mathscr{A}_x$
consisting from of almost periodic functions. Let us denote
\begin{eqnarray}
\begin{array}{l}
\mathscr{P}_y^{\beta} = \left\{  \phi_{x(t)}^{\beta}(t) :~ x(t)
\in \mathscr{P}_x  \right\}\subset \mathscr{A}_y^{\beta}.
\end{array}
\end{eqnarray}
Condition $(C4)$ implies that there is a one-to-one correspondence
between the sets $\mathscr{P}_x$ and $\mathscr{P}_y^{\beta}.$
Therefore, $\mathscr{P}_y^{\beta}$ is also infinite countable.
Furthermore, Lemma \ref{almost} implies that
$\mathscr{P}_y^{\beta}$ consists from almost periodic functions.
Hence the point (i) in the definition of Li-Yorke chaotic set for
$\mathscr{A}_y^{\beta}$ is verified.

Now, suppose that $\mathscr{C}_x$ is an uncountable scrambled subset of $\mathscr{A}_x$. Let us introduce
\begin{eqnarray}
\begin{array}{l}
\mathscr{C}_y^{\beta} = \left\{  \phi_{x(t)}^{\beta}(t) :~ x(t) \in \mathscr{C}_x  \right\}.
\end{array}
\end{eqnarray}
Condition $(C4)$ again implies that there is a one-to-one correspondence between the sets $\mathscr{C}_x$ and $\mathscr{C}_y^{\beta}.$ Therefore, $\mathscr{C}_y^{\beta}$ is also uncountable. Under the same condition, it is easy to verify that there does not exist any almost periodic function inside the set $\mathscr{C}_y^{\beta}.$ Indeed, assume that $y\in\mathscr{C}_y^{\beta}$ is almost periodic. Then by \eqref{ess1}, $y(t)$ is uniformly continuous on $\mathbb{R}$. Furthermore, using conditions (C3) and (C5), we see that $x(t)$ and $\nu_\beta(t)$ are uniformly continuous on $\mathbb{R}$ as well. So assumption (C1) together with \eqref{e3} give that $y'(t)$ is also uniformly continuous on $\mathbb{R}$. Then by \cite{Lev82}, $y'(t)$ is almost periodic, hence by \cite{Hale80}, function
$$
z(t)=y'(t)-f(y(t)+\nu_\beta(t),t)+f(\nu_\beta(t),t)
$$
is almost periodic. Next, condition (C4) implies that $h : \mathbb{R}^n\to \mathbb{R}^n$ is a homeomorphism (see \cite[Theorems (2.7.1), (5.1.4), (5.4.11)]{Ber77}). So by \eqref{e3} we obtain $x(t)=h^{-1}(z(t))$, then by \cite{Lev82}, $x(t)$ is almost periodic which contradicts to
$x\in \mathscr{C}_x$. So any $y\in\mathscr{C}_y^{\beta}$ is not almost periodic. As a matter of fact we have shown that $x\in \mathscr{C}_x$ is almost periodic if and only if the corresponding $y\in\mathscr{C}_y^{\beta}$ is almost periodic.

Next, since the collection $\mathscr{A}_x$  is assumed to be
chaotic in the sense of Li-Yorke, each couple of functions inside
$\mathscr{C}_x \times  \mathscr{C}_x $ is proximal. Thus, Lemma
\ref{proximality_proof} implies that the same is valid for each
couple inside $\mathscr{C}_y^{\beta} \times \mathscr{C}_y^{\beta}
.$ Hence the point (ii) in the definition of Li-Yorke chaotic set
for $\mathscr{A}_y^{\beta}$ is verified as well.

On the other hand, Lemma $\ref{seperation_proof}$ implies the
existence of positive real numbers $\epsilon_1$ and
$\overline{\Delta}$ such that each couple of functions
$\left(y(t),\overline{y}(t) \right) \in \mathscr{C}_y^{\beta}
\times  \mathscr{C}_y^{\beta} $ are frequently
$\left(\epsilon_1,\overline{\Delta}\right)-$separated. The same
property is true also for each couple of sequences inside $
\left(\mathscr{C}_y^{\beta}  \times \mathscr{G}_y^\beta\right)$
for the set $\mathscr{G}_y^\beta$ of all almost periodic functions
in $\mathscr{C}_y^{\beta}$. This verifies the point (iii) in the
definition of Li-York chaotic set for $\mathscr{A}_y^{\beta}$.
Consequently, the set of functions $\mathscr{A}_y^{\beta}$ is a
Li-Yorke chaotic set. The proof is finalized.
$\square$

\begin{remark}\label{alternat}
When we consider ``$i\bar l$-periodic for some $i\in\mathbb{N}$" in
the definitions of scrambled set $\mathscr{C}_x$ and Li-Yorke
chaotic set $\mathscr{A}_x$ instead of ``almost periodic", then by
using Lemma \ref{almost}, we see that Theorem
\ref{li-yorke_theorem} is valid also for this alternative periodic
case by considering any periodic $\nu_\beta(t)$. Roughly writing,
now scrambled set $\mathscr{C}_x$ contains functions nonresonant
with the unperturbed, i.e., without $h(x)$, part of \eqref{e3} for
any periodic $\nu_\beta(t)$. On the other hand, the set in the
part (i) of definition of Li-York chaotic set contains resonant
functions.
\end{remark}

\begin{remark}
System \eqref{e1} may possess bounded solutions other than $\nu_{\beta}(t)$, $ \beta \in S_m.$  Then there may exist  a replicated chaos  corresponding  for  each  of such solution, but  verification of that  is a difficult task in general, which would need  additional  assumptions for the system. This is why  we  are  satisfied with  the  proof of the  chaos   around  an almost periodic (or just periodic) solution $\nu_\beta(t)$ for $\beta \in S_m$.
\end{remark}

\section{An Example}

This part of the paper is devoted to an illustrative example. First of all, we will take into account a forced Duffing equation, which is known to be chaotic in the sense of Li-Yorke, as the source of chaotic perturbations. The forcing term in this equation will be in the form of a relay function to ensure the presence of Li-Yorke chaos. Detailed theoretical as well as numerical results concerning relay systems can be found in the papers \cite{Akh1,Akh4,Akh5,Akh7,Akh8}. In order to provide the replication of chaos, we will perturb another Duffing equation, which admits a homoclinic orbit, by the solutions of the former.

Another issue that we will focus on is the stabilization of unstable quasi-periodic motions. In the literature, control of chaos is understood as the stabilization of unstable periodic orbits embedded in a chaotic attractor \cite{Gon04,Sch99}. However, in this section, we will demonstrate the stabilization quasi-periodic motions instead of periodic ones, and this is one of the distinguishing features of our results. The existence of unstable quasi-periodic motions embedded in the chaotic attractor will be revealed by means of an appropriate chaos control technique based on the Ott-Grebogi-Yorke (OGY) \cite{Ott90} and Pyragas \cite{Pyragas92} control methods.

Let us consider the following forced Duffing equation,
\begin{equation}\label{ex1}
x''+0.82x'+1.4x+0.01x^{3}=v(t,\zeta,\lambda),
\end{equation}
where the relay function $v(t,\zeta,\lambda)$ is defined as
\begin{equation} \label{relay}
v(t,\zeta,\lambda)= \begin{cases}
0.3,\  \textrm{if} \ \zeta_{2j}(\lambda) < t \le \zeta_{2j+1}(\lambda), \ j\in \mathbb{Z}, \\
1.9,\ \textrm{if} \  \zeta_{2j-1}(\lambda) < t \le \zeta_{2j}(\lambda), \ j\in \mathbb{Z}.
\end{cases}
\end{equation}
In (\ref{relay}), the sequence $\zeta=\left\{\zeta_j\right\}_{j\in \mathbb Z},$ $\zeta_0 \in[0,1],$ of switching moments is defined through the equation $\zeta_{j}  = j + \kappa_j,$ $j\in \mathbb Z,$ and the sequence $\left\{\kappa_j\right\}_{j\in \mathbb Z}$ is a solution of the logistic map
\begin{eqnarray}\label{logistic_map}
\kappa_{j+1} = \lambda \kappa_j (1-\kappa_j).
\end{eqnarray}

The interval $[0,1]$ is invariant under the iterations of (\ref{logistic_map}) for the values of $\lambda$ between $1$ and $4$, and the map possesses Li-Yorke chaos for $\lambda = 3.9$ \cite{Li75}.

Making use of the new variables $x_1=x$ and $x_2=x',$ one can reduce equation (\ref{ex1}) to the system
\begin{equation}\label{ex2}
\begin{array}{l}
x'_1 = x_2 \\
x'_2 = -1.4 x_1 - 0.82 x_2 - 0.01 x_1^3 + v(t,\zeta,\lambda).
\end{array}
\end{equation}
According to the results of \cite{Akh1}, for each $\zeta_0 \in [0,1],$ system (\ref{ex2}) with $\lambda =3.9$ possesses a bounded on $\mathbb R$ solution, and the collection $\mathscr{A}_x$ consisting of all such bounded on $\mathbb R$ solutions is Li-Yorke chaotic. Moreover, for each natural number $m$, system (\ref{ex2}) admits unstable periodic solutions with periods $2m.$ The reader is referred to \cite{Akh1,Akh4,Akh5,Akh7,Akh8,AkhF} for more information about the dynamics of relay systems.

\begin{figure*}[ht]
\centering
\epsfig{file=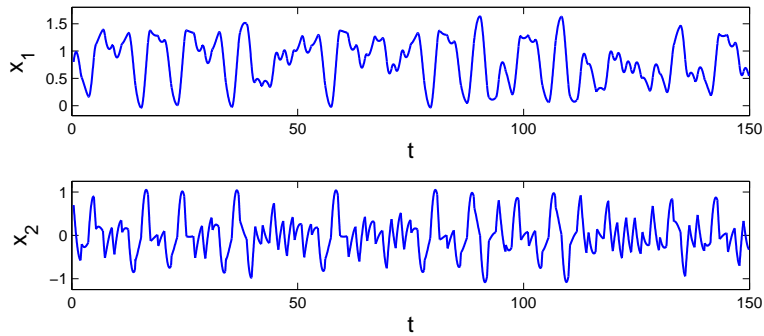, width=11.5cm}
\caption{The chaotic behavior of system (\ref{ex2}).}
\label{fig1}
\end{figure*}

Figure \ref{fig1} shows the solution of system (\ref{ex2}) with $\lambda=3.9,$ $\zeta_0=0.41$ corresponding to the initial data $x_1(0.41)=0.8,$ $x_2(0.41)=0.7.$ The simulation results seen in Figure \ref{fig1} confirm the presence of chaos in (\ref{ex2}).

Next, let us consider the following Duffing equation \cite{Awrejcewicz04},
\begin{eqnarray}\label{ex3}
z''+0.15 z' -0.5 z (1-z^2)=0.2\sin(0.9 t).
\end{eqnarray}
It was mentioned in \cite{Awrejcewicz04} that the equation (\ref{ex3}) is chaotic, and it admits a homoclinic orbit.

By means of the variables $z_1=z$ and $z_2=z',$ equation (\ref{ex3}) can be written as a system in the following form,
\begin{eqnarray}\label{ex4}
\begin{array}{l}
z'_1 =z_2 \\
z'_2  = - 0.15 z_2 +0.5 z_1(1-z_1^2)+0.2\sin(0.9 t).
\end{array}
\end{eqnarray}
We perturb system  (\ref{ex4}) with the solutions of (\ref{ex2}) to obtain the system
\begin{eqnarray}\label{ex5}
\begin{array}{l}
u'_1 =u_2  + 1.2 (x_1 +0.1\sin(x_1))\\
u'_2  = - 0.15 u_2 +0.5 u_1  (1-u_1^2)   +0.4 \arctan(x_2)+0.2\sin(0.9 t).
\end{array}
\end{eqnarray}

According to Theorem \ref{li-yorke_theorem}, system (\ref{ex5}) possesses chaos in the sense of Li-Yorke. Moreover, since the period of the function $0.2\sin(0.9 t)$ and the periods $2m,$ $m\in \mathbb N,$ of the unstable periodic motions of (\ref{ex2}) are incommensurate, there are infinitely many unstable quasi-periodic motions embedded in the chaotic attractor of (\ref{ex5}).

\begin{figure*}[ht]
\centering
\epsfig{file=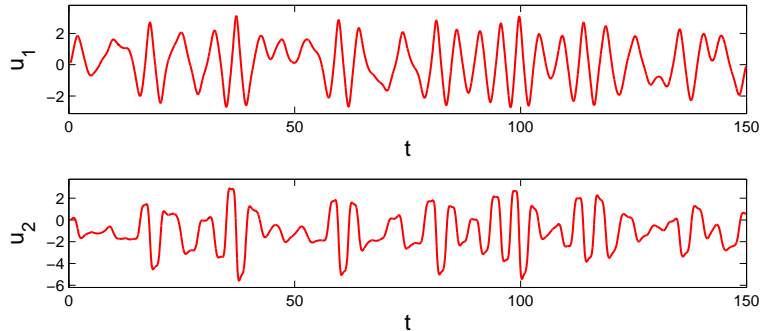, width=11.5cm}
\caption{The chaotic solution of the perturbed system (\ref{ex5}) with $u_1(0.41)=0.12$ and $u_2(0.41)=0.013.$ The solution $(x_1(t),x_2(t))$ of (\ref{ex2}), which is represented in Figure \ref{fig1}, is used as the perturbation in (\ref{ex5}).}
\label{fig2}
\end{figure*}

In system (\ref{ex5}), as the perturbation, we use the solution $(x_1(t),x_2(t))$ of (\ref{ex2}) which is represented in Figure \ref{fig1}, and depict in Figure \ref{fig2} the solution of (\ref{ex5}) with the initial data $u_1(0.41)=0.12 $ and $u_2(0.41)=0.013.$ The simulations seen in Figure \ref{fig2} support the result of Theorem \ref{li-yorke_theorem} such that the perturbed system (\ref{ex5}) exhibits Li-Yorke chaos.

\begin{figure}[H]
\centering
\epsfig{file=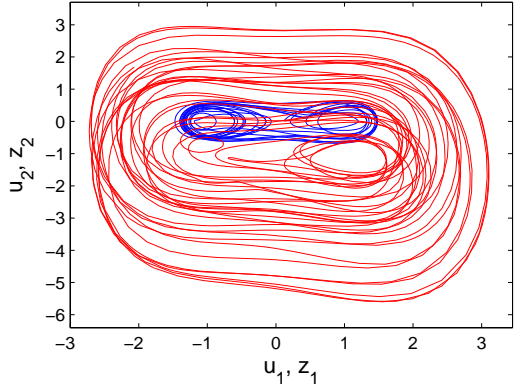, width=7.5cm}
\caption{Chaotic trajectories of systems (\ref{ex4}) and (\ref{ex5}). The trajectory of (\ref{ex4}) with $z_1(0.41)=0.12,$ $z_2(0.41)=0.013$ is represented in blue color, while the trajectory of (\ref{ex5}) corresponding to $u_1(0.41)=0.12,$ $u_2(0.41)=0.013$ is shown in red color. One can observe that the systems (\ref{ex4}) and (\ref{ex5}) admit different chaotic attractors.}
\label{fig3}
\end{figure}

In order to verify that the unperturbed system (\ref{ex4}) and the perturbed system (\ref{ex5}) possess different chaotic attractors, we depict in Figure \ref{fig3} the trajectories of (\ref{ex4}) and (\ref{ex5}) corresponding to the initial data $z_1(0.41)=0.12,$ $z_2(0.41)=0.013$ and $u_1(0.41)=0.12,$ $u_2(0.41)=0.013,$ respectively. Here, the trajectory of (\ref{ex4}) is depicted in blue color and the trajectory of (\ref{ex5}) is shown in red color. It is seen in Figure \ref{fig3} that even if the same initial data is used, systems (\ref{ex4}) and (\ref{ex5}) generate completely different chaotic trajectories. In other words, the applied perturbation annihilates the chaotic attractor of (\ref{ex4}) and causes a new one to be formed in the dynamics of (\ref{ex5}).

Now, we will confirm the presence of quasi-periodic motions embedded in the chaotic attractor of (\ref{ex5}) by stabilizing one of them through a control technique based on the OGY \cite{Ott90} and Pyragas \cite{Pyragas92} methods. The idea of the control procedure depends on the usage of both the OGY control for the discrete-time dynamics of the logistic map (\ref{logistic_map}), which the source of chaotic motions in the forced Duffing equation (\ref{ex2}), and the Pyragas control for the continuous-time dynamics of (\ref{ex5}). The simultaneous usage of both methods will give rise to the stabilization of a quasi-periodic solution of (\ref{ex5}) since (\ref{ex2}) and (\ref{ex4}) admit periodic motions with incommensurate periods.

Let us explain briefly the OGY control method for the map (\ref{logistic_map}) \cite{Sch99}. Suppose that the parameter $\lambda$ in the  map (\ref{logistic_map}) is allowed to vary in the range $[3.9-\varepsilon, 3.9+\varepsilon]$, where $\varepsilon$ is a given small positive number. Consider an arbitrary solution $\left\{\kappa_j\right\},$ $\kappa_0\in[0,1],$ of the map and denote by $\kappa^{(i)},$ $i=1,2,\ldots ,p,$ the target $p-$periodic orbit to be stabilized.
In the OGY control method \cite{Sch99}, at each iteration step $j$ after the control mechanism is switched on, we consider the logistic map with the parameter value $\lambda=\bar \lambda_j,$ where
\begin{eqnarray}\label{control}
\bar \lambda_j=3.9 \left(1+\frac{(2\kappa^{(i)}-1)(\kappa_{j}-\kappa^{(i)})}{\kappa^{(i)}(1-\kappa^{(i)})} \right),
\end{eqnarray}
provided that the number on the right hand side of the formula $(\ref{control})$ belongs to the interval $[3.9-\varepsilon, 3.9+\varepsilon].$ In other words, formula (\ref{control}) is valid if the trajectory $\left\{\kappa_j\right\}$ is sufficiently close to the target periodic orbit. Otherwise, we take $\bar \lambda_{j}=3.9,$ so that the system evolves at its original parameter value, and  wait until the trajectory $\left\{\kappa_j\right\}$ enters in a sufficiently small neighborhood of the periodic orbit $\kappa^{(i)},$ $i=1,2,\ldots, p,$ such that the inequality $-\varepsilon \le 3.9 \displaystyle\frac{(2\kappa^{(i)}-1)(\kappa_{j}-\kappa^{(i)})}{\zeta^{(i)}(1-\zeta^{(i)})} \le \varepsilon$ holds. If this is the case, the control of chaos is not achieved immediately after switching on the control mechanism. Instead, there is a transition time before the desired periodic orbit is stabilized. The transition time increases if the number $\varepsilon$ decreases \cite{Gon04}.

On the other hand, according to the Pyragas control method \cite{Gon04,Pyragas92}, an unstable periodic solution with period $\tau_0$ can be stabilized by using an external perturbation of the form $C[s(t-\tau_0)-s(t)],$ where $C$ is the strength of the perturbation, $s(t)$ is a scalar signal which is given by some function of the state of the system and $s(t-\tau_0)$ is the signal measured with a time delay equal to $\tau_0.$

To stabilize an unstable quasi-periodic solution of (\ref{ex5}), we construct the following control system,
\begin{equation}\label{ex6}
\begin{array}{l}
w'_1 = w_2 \\
w'_2 = -1.4 w_1 - 0.82 w_2 - 0.01 w_1^3 + v(t,\zeta,\bar \lambda_j) \\
w'_3 =w_4  + 1.2 (w_1 +0.1\sin(w_1))\\
w'_4  = - 0.15 w_4 +0.5 w_3  (1-w_3^2) +0.4 \arctan(w_2) \\ +0.2\sin(0.9 w_5) + C[w_4(t-2\pi/0.9)-w_4(t)]\\
w'_5=1.
\end{array}
\end{equation}
\begin{figure*}[ht]
\centering
\epsfig{file=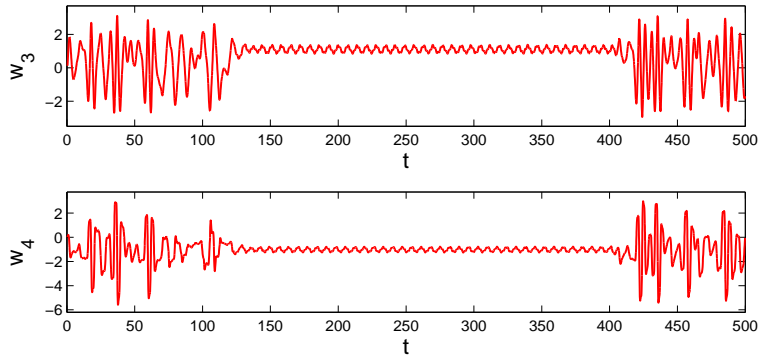, width=11.5cm}
\caption{Chaos control of system (\ref{ex5}). We make use of the OGY control method around the fixed point $2.9/3.9$ of the logistic map (\ref{logistic_map}). The control is switched on at $t=\zeta_{70}.$ The OGY control switched off at $t=\zeta_{350},$ while the Pyragas control is switched off at $t=\zeta_{400}.$ The values $\varepsilon=0.085$ and $C=2.6$ are used in the simulation.}
\label{fig4}
\end{figure*}
\begin{figure*}[ht]
\centering
\includegraphics[width=11.5cm]{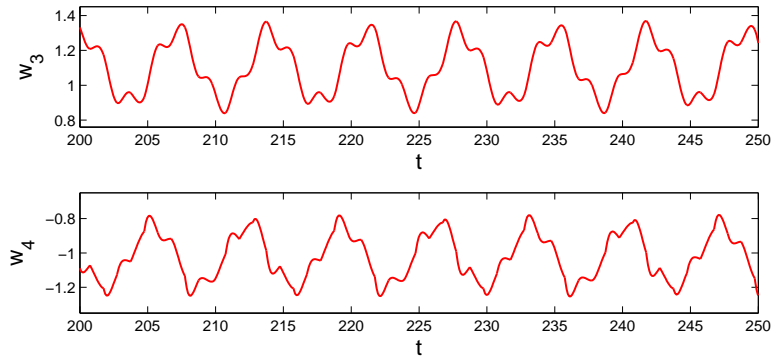}
\caption{The stabilized quasi-periodic solution of system (\ref{ex5}).}
\label{fig5}
\end{figure*}
We make use of the OGY control method around the fixed point $2.9/3.9$ of the logistic map (\ref{logistic_map}) so that $\bar \lambda_j$ in (\ref{ex6}) is given by the formula (\ref{control}) in which $\kappa^{(i)}\equiv 2.9/3.9$. The control mechanism is switched on by using the values $\varepsilon=0.085$ and $C=2.6.$ The OGY control switched off at $t=\zeta_{350}$ and the Pyragas control is switched off at $t=\zeta_{400}.$ In other words, we take $\lambda_j=3.9$ after $t=\zeta_{350},$ and take $C=0$ after $t=\zeta_{400}.$ Figure \ref{fig4} shows the simulation results for the $w_3$ and $w_4$ coordinates of system (\ref{ex6}) corresponding to the initial data $w_1(0.41)=0.8,$ $w_2(0.41)=0.7,$ $w_3(0.41)= 0.12,$ $w_4(0.41)=0.013,$ $w_5(0.41)=0.41.$ It is seen in the figure that the quasi-periodic solution of (\ref{ex5}) is stabilized. To present a better visuality, the stabilized quasi-periodic solution of (\ref{ex5}) is shown in Figure \ref{fig5} for $200\le t \le 250.$

\section{Appendix}

For the reader convenience, we present and prove a Growall-Coppel type inequality (see \cite{BS}) result used in this paper.

\begin{theorem}\label{gr} Let $a,b,c$ be nonnegative constants, $\gamma>0$ and $u\in C([p,q],\mathbb{R})$ be nonnegative on an interval $[p,q]$, $p<q$, satisfying
$$
u(t)\le a+b\left(e^{-\gamma(t-p)}+e^{-\gamma(q-t)}\right)+c\int_p^qe^{-\gamma|t-s|}u(s)ds
$$
on $[p,q]$. If $2c<\gamma$ then
$$
u(t)\le \frac{a\gamma}{\gamma-2c}+\frac{b}{c}(\gamma-\delta)\left(e^{-\delta(t-p)}+e^{-\delta(q-t)}\right)
$$
for any $t\in[p,q]$, where $\delta=\sqrt{\gamma^2-2c\gamma}$.
\end{theorem}
\noindent \textbf{Proof.}
By \cite[Theorems 2.3, 2.4]{BS}, functions
\begin{eqnarray*}
u_1(t)&=&\frac{b}{c}(\gamma-\delta)e^{-\delta(t-p)},\quad t\ge p,\\
u_2(t)&=&\frac{b}{c}(\gamma-\delta)e^{-\delta(q-t)},\quad t\le q,\\
u_3(t)&=&\frac{a\gamma}{\gamma-2c}
\end{eqnarray*}
satisfy the equations
\begin{eqnarray*}
u_1(t)&=&be^{-\gamma(t-p)}+c\int_{p}^\infty e^{-\gamma|t-s|}u_1(s)ds,\\
u_2(t)&=&be^{-\gamma(q-t)}+c\int_{-\infty}^qe^{-\gamma|t-s|}u_2(s)ds,\\
u_3(t)&=&a+c\int_{-\infty}^\infty e^{-\gamma|t-s|}u_3(s)ds,
\end{eqnarray*}
respectively. Since all $u_1(t),u_2(t),u_3(t)$ are nonnegative, the function
$$
u_4(t)=u_1(t)+u_2(t)+u_3(t)
$$
satisfies
\begin{eqnarray*}
u_4(t) \ge a+b\left(e^{-\gamma(t-p)}+e^{-\gamma(q-t)}\right) + \displaystyle c\int_p^qe^{-\gamma|t-s|}u_4(s)ds
\end{eqnarray*}
on $[p,q]$. Next, consider the operator $\Upsilon : C([p,q],\mathbb{R})\to C([p,q],\mathbb{R})$ given by
\begin{eqnarray*}
 \Upsilon u(t)  =  a+b\left(e^{-\gamma(t-p)}+e^{-\gamma(q-t)}\right) + \displaystyle c\int_p^qe^{-\gamma|t-s|}u(s)ds.
\end{eqnarray*}
Then it is nondecreasing and by \cite[p. 14]{BS}, it is contractive. So it has a unique fixed point $u_*\in C([p,q],\mathbb{R})$, i.e., $u_*=\Upsilon u_*$. Since $u\le \Upsilon u$ and $u_4\ge \Upsilon u_4$, by standard arguments (see \cite[Theorem 2.2]{BS}), we get $u\le u_*\le u_4$. The proof is finished.
$\square$

\section*{Acknowledgements}

This work was supported by the Grants VEGA-MS
1/0071/14, VEGA-SAV 2/0153/16, by the Slovak Research and Development
Agency under the contract No. APVV-14-0378, and by the National Scholarship
Programme of the Slovak  Republic.


\end{document}